\newcommand{\text}{\rm }

\input{aipcheck}

\documentclass[
    ,final            % use final for the camera ready runs
  ]
  {aipproc}

\layoutstyle{8x11single}

\begin{document}

\title{Why mean {\em p}$_{\rm \bf T}$ is interesting}


\author{Michal Praszalowicz}{
  address={M. Smoluchowski Institute of Physics, Jagiellonian University, \\
S. Lojasiewicz str. 11, 30-348 Krakow, Poland.}
}

%\author{Larry McLerran}{
%  address={Physics Dept, Bdg. 510A, Brookhaven National Laboratory, Upton, NY-11973, USA,\\
%RIKEN BNL Research Center, Bldg. 510A, Brookhaven National Laboratory, Upton, NY 11973, USA,\\
%Physics Department, China Central Normal University, Wuhan, 430079, China.}
%}

%\author{<author3>}{
%  address={<common address for author2 and author3>}
%  ,altaddress={<author1 address>} % additional visiting address
%}

\begin{abstract}
We discuss energy
dependence of mean $p_{\rm T}$ correlation with $N_{\rm ch}$ basing on general
features of high energy collisions such as saturation and geometrical scaling.
We use Color Glass Condensate calculation of an effective interaction
radius that scales as a third root of multiplicity, and then saturates. With
this model input we construct scaling variable for $\langle  p_{\rm T} \rangle (N_{\rm ch})$
at different energies  both for pp and pPb collisions, and show that recent ALICE data
indeed does exhibit this scaling property. We discuss  energy dependence of the interaction
radius and argue that since the radius cannot grow too large, a universal behavior of
$\langle  p_{\rm T} \rangle$ for large multiplicities is expected.

\end{abstract}

\maketitle

%%%%%%%%%%%%%%%%%%%%%%%%%%%%%%%%%%%%%%%%%%%%
%% MAINMATTER
%%%%%%%%%%%%%%%%%%%%%%%%%%%%%%%%%%%%%%%%%%%%

%\section{<A section>}

Correlations are always interesting since they are sensitive to the fine details of interactions.
For example standard PYTHIA Monte Carlo fails in the case of $\left\langle p_{\rm {T}}\right\rangle $ 
correlation with $N_{\rm ch}$ (see {\em e.g.} Refs.~\cite{Abelev:2013ala,Abelev:2013bla}), 
while it does describe well one particle spectra and total multiplicities.
New effect called color recombination (see {\em e.g.} \cite{Gustafson:2008zz})
has to be added to PYTHIA to take care of rather strong rise 
of  $\left\langle p_{\rm {T}}\right\rangle $ with $N_{\rm ch}$. On the other hand newer MC generator 
EPOS \cite{Pierog:2013ria} that has saturation effects built in does not require special tuning to describe 
$\left\langle p_{\rm {T}}\right\rangle $ in function of $N_{\rm ch}$ \cite{Abelev:2013ala,Abelev:2013bla}. 
In this talk that is based
on Refs.~\cite{McLerran:2013oju,McLerran:2014apa} 
(and where a complete list of references can be found) we shall study
consequences of saturation and geometrical scaling (GS) for mean transverse momenta at the LHC.
Here the energy must be really large since the character of 
 $\left\langle p_{\rm {T}}\right\rangle $ dependence on $N_{\rm ch}$ changes dramatically 
 (see Fig.~1 in Ref.~\cite{Armesto:2008zz})
 from
 the ISR energies (where it decreases or stays constant) to the LHC energies (where it rises). 
 Theoretical interest in $\left\langle p_{\rm {T}}\right\rangle $  correlation with $N_{\rm ch}$
 goes back to the paper by L. van Hove \cite{VanHove:1982vk} where he pointed out 
 that qualitative change is expected
 in the presence of the QCD phase transition. Today this motivation is perhaps less important
 since the QCD deconfining transition is believed to be a soft crossover.
 
 From the saturation point of view high energy central rapidity production of moderate $p_{\rm T}$
 particles can be viewed as a result of a collision of two gluonic clouds characterized by 
 one saturation scale that depends on gluon longitudinal momenta $x_1 \sim x_2$ denoted in the
 following as $x$:
 \begin{equation}
Q_{\rm {s}}^{2}(x)=Q_{0}^{2} \left(  \frac{x_{0}}{x} \right)  ^{\lambda
}\label{Qsdef}%
\end{equation}
where $Q_{0}$ is an arbitrary scale parameter for
which we take 1 ~GeV$/c$, and for $x_0$ we take $10^{-3}$. 

An immediate consequence of (\ref{Qsdef}) is GS of particle (or strictly
speaking gluon) spectra \cite{McLerran:2010ex}:
\begin{equation}
\frac{dN_{\rm g}}{dyd^{2}p_{\bf {T}}}=S_{\bot}\mathcal{F}(\tau)\label{dNdydpT}%
\end{equation}
where%
\begin{equation}
\tau=\frac{p_{\bf {T}}^{2}}{Q_{\bf {s}}^{2}(x)}%
\end{equation}
is the scaling variable. Here $S_{\bot}$ is a transverse area which will be specified later.
Logarithmic corrections due the running of the strong coupling constant are neglected in 
Eq.~(\ref{dNdydpT}).

In order to integrate (\ref{dNdydpT}) over $d^2 p_{\rm T}$ we have to change integration
variable to $d \tau$ which gives
\begin{equation}
\frac{dN_{\rm g}}{dy}=A\,S_{\bot}\bar{Q}_{\mathrm{s}}^{2}(W)\label{dNoverdy}%
\end{equation}
where $A$ is an energy independent integral of the universal function $\mathcal{F}(\tau)$
over $d \tau$.
Here 
\begin{equation}
\bar{Q}_{\mathrm{s}}(W)=Q_{0}\left(  \frac{W}{Q_{0}}\right)  ^{\lambda
/(2+\lambda)}
\label{Qbars}%
\end{equation}
is an {\sl  average} saturation scale which gives 
(provided that $S_\bot$ is energy {\sl independent})
power like growth of multiplicity
with the scattering energy $W$
 -- a fact well confirmed by recent LHC data on multiparticle
production.

\begin{figure}[h!]
  \includegraphics[height=.3\textheight]{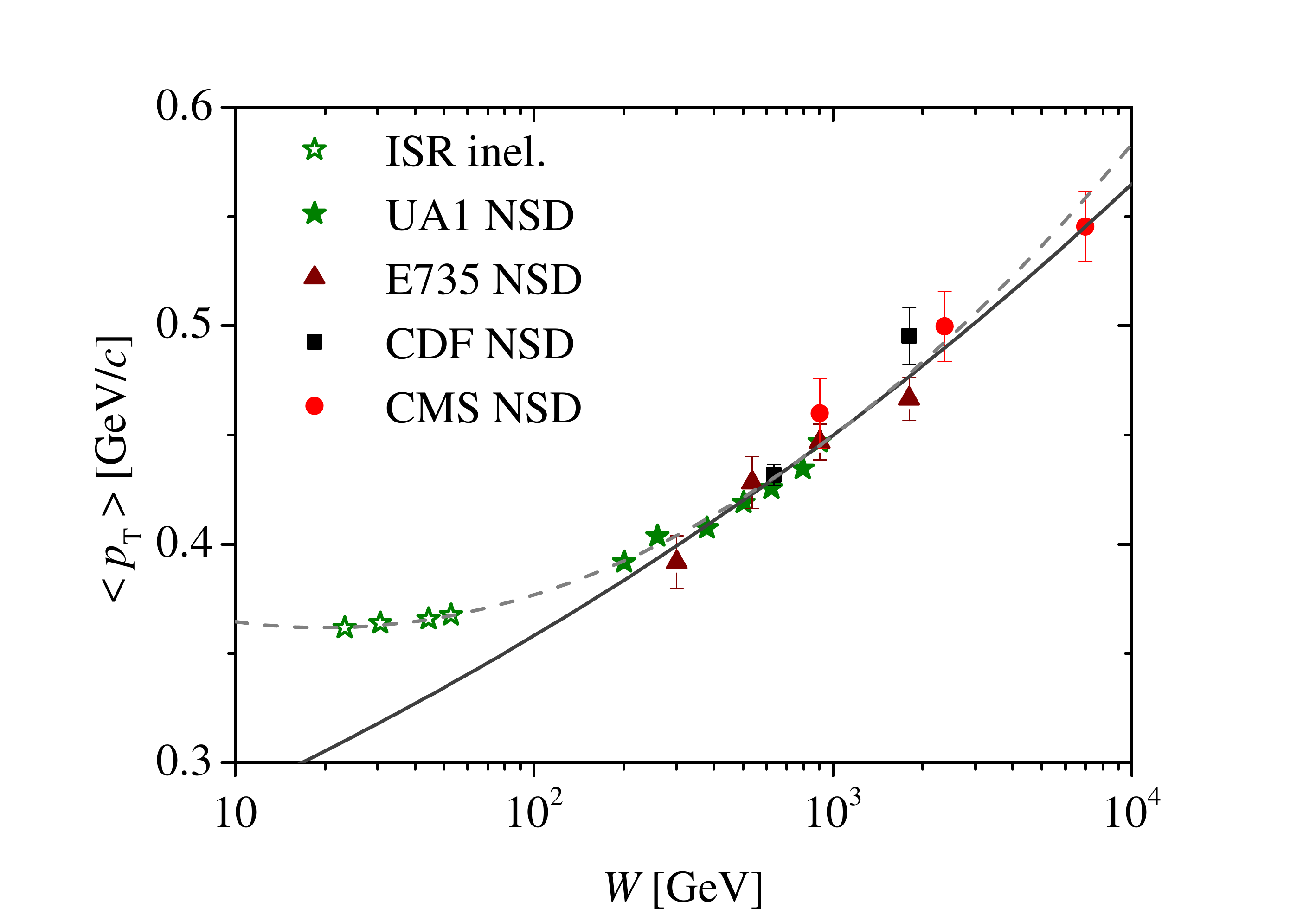}
  \caption{Mean transverse momentum as a function of energy with the power-like
  fit $0.227\times W^{0.099}$ (solid line). Compilation of data taken from Ref.~\cite{Khachatryan:2010us}.  
  The dashed line corresponds to the CMS logarithmic fit aimed at describing also low energy data: $0.413-0.0171\,\ln s +0.00143\, \ln^2 s$, with $W^2=s$. (Figure from Ref.~\cite{McLerran:2014apa}.)}%
\label{fig:meanpTW}
\end{figure}

An immediate consequence of Eq.~(\ref{dNoverdy}) is that \cite{McLerran:2014apa}
\begin{equation}
\left\langle p_{\rm {T}}\right\rangle  \sim \bar{Q}_{\mathrm{s}}(W)\label{meanpT1}\, ,%
\end{equation}
which means that $\left\langle p_{\rm {T}}\right\rangle$ rises with energy as $W^{\lambda
/(2+\lambda)}$. 
In order to find numerical value of $\lambda$ we have performed simple
analysis \cite{McLerran:2014apa} based on GS hypothesis, requiring that ratios
\begin{equation}
\mathcal{R}_{ik}(\tau)=\frac{dN(W_{i},\tau)}{dyd^{2}p_{\rm {T}}}/\frac
{dN(W_{k},\tau)}{dyd^{2}p_{\rm {T}}}\label{Rik}%
\end{equation}
are as close to unity as possible over the widest range of $\tau$, with the result $\lambda \approx 0.22$.
As a consequence mean $p_{\rm T}$ rises with energy as a power of energy $W$, which, 
as shown in Fig.~\ref{fig:meanpTW},
is in perfect agreement with high energy data \cite{Khachatryan:2010us}.

Replacing $\bar{Q}_{\mathrm{s}}(W)$ in Eq.~(\ref{fig:meanpTW}) by (\ref{dNoverdy}) 
at fixed $W=W_0$
and adding
a constant piece, which takes into account nonperturbative effects
and contributions from the particle masses, we arrive at \cite{McLerran:2013oju}:%
\begin{equation}
\left.  \left\langle p_{\text{T}}\right\rangle \right\vert _{W_0}%
=\alpha+\beta\,\sqrt{\frac{\pi \, N_{\text{ch}}}{\left.  S_{\bot}(\sqrt[3]{\gamma
N_{\text{ch}}})\right\vert _{W_0}}}\, .\label{meanpTfit}%
\end{equation}
Here $\alpha$ and $\beta$ are constants that do not depend on energy and $\gamma$ is a parameter
that, using parton-hadron duality assumption, relates number of gluons to the number of charged particles.

Three important comments concerning Eq.~(\ref{meanpTfit}) are in order.
First, transverse size entering Eqs.~(\ref{dNdydpT}) and (\ref{dNoverdy})
in a situation where we fix number of particles
corresponds to the overlap between two hadrons colliding at given impact parameter $b$.
As such it does depend on the multiplicity itself. Second, in Eq.~(\ref{meanpTfit})
we have assumed explicit relation between $S_\bot$ and $N_g = \gamma N_{\rm ch}$ that has been calculated
in the CGC effective theory \cite{Bzdak:2013zma}, which essentially says that particle multiplicity 
is proportional
to the interaction {\sl volume}. Explicitly: it rises with volume linearly up to some
maximal $V_{\rm max}$ and then at fixed volume it rises further on only due to
fluctuations. This allows  to define interaction {\sl radius} $R \sim S_\bot^{1/2}$
whose explicit dependence on $N_g$ can be found in Refs.~\cite{McLerran:2013oju,McLerran:2014apa}. 
And finally,
the CGC calculations have been performed for fixed scattering energy $W_0$,
so in principle in order to find mean $p_{\rm T}$ at a different energy one should
recalculate $S_\bot$ in Eq.~(\ref{meanpTfit}). In the following we shall argue that
this is essentially not necessary. 

Indeed, although after Eq.~(\ref{Qbars}) we have argued that $S_\bot$ does not
depend on energy, in a situation when we change energy at {\sl fixed} multiplicity, we
have to vary $S_\bot$ accordingly, so that Eq.~(\ref{dNoverdy}) remains satisfied.  
As a result at higher energies the same
number of particles comes from the smaller overlap, \emph{i.e.} from larger $b$.  
Taking into account this {\sl induced} energy dependence of the interaction radius
\begin{equation}
\left. R \right\vert _{W} = \left(  \frac{W_0}{W}\right)^{\lambda/(2+\lambda)}
\left. R \right\vert _{W_0}
\label{RWdep}
\end{equation} 
we arrive at
\begin{equation}
\left.  \left\langle p_{\text{T}}\right\rangle \right\vert _{W}=
{\alpha}
+{\beta}\,\frac{\sqrt{N_{\text{ch}}}}{\left.  R(\sqrt[3]{{\gamma} N_{\text{ch}}%
})\right\vert _{W}}=
{\alpha}+{\beta}\,\left(  \frac{W}{W_{0}}\right)
^{\lambda/(2+\lambda)}\frac{\sqrt{N_{\text{ch}}}}{\left.  R(\sqrt[3]{{\gamma}
N_{\text{ch}}})\right\vert _{W_{0}}}\label{pTWdep}%
\end{equation}
where $W_{0}$ corresponds to the energy for which the interaction radius has
been computed in some explicit model. In the CGC theory that we use here
\cite{Bzdak:2013zma}
 $W_0 = 7$~TeV
for pp and $5.02$~TeV for pPb.

\begin{figure}[h]
\centering
\includegraphics[height=.25\textheight]{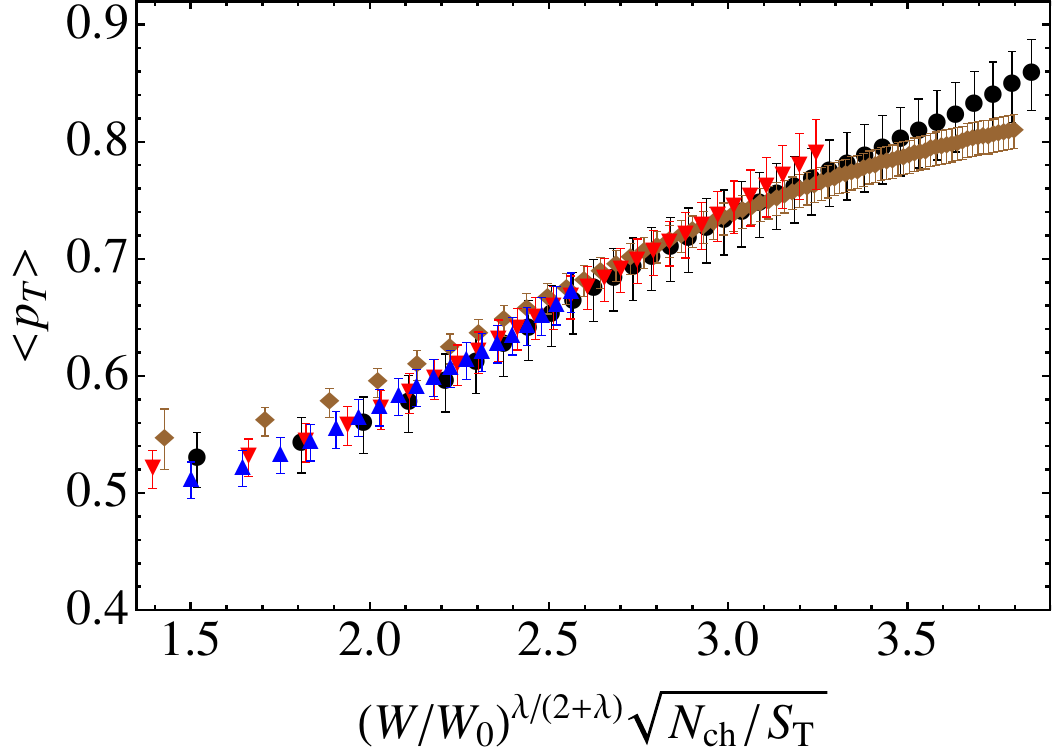}\caption{Mean
$\left\langle p_{\rm {T}}\right\rangle $ in p-p collisions at 7 TeV (full
black circles), 2.76 TeV (full red down-triangles), 0.9 TeV (full blue
up-triangles) and in p-Pb collisions at 5.02 TeV (full brown diamonds) plotted
in terms of scaling variable $(W/W_{0})^{\lambda/(2+\lambda)}\sqrt
{N_{\mathrm{ch}}/ S_\bot}$. For p-p $W_{0}=7$~TeV and for p-Pb $W_{0}%
=5.02$~TeV. (Figure from Ref.~\cite{McLerran:2014apa}.)}%
\label{fig:scaled}%
\end{figure}

It follows from Eq.~(\ref{pTWdep}) that there exists a universal scaling variable 
\[
\left( W/W_0 \right)^{\lambda/(2+\lambda)}\sqrt{N_{\text{ch}}}/{\left.  R(\sqrt[3]{{\gamma}
N_{\rm {ch}}})\right\vert _{W_{0}}}
\]
for which all mean $p_{\rm T}$ data should fall on one curve. This scaling property is clearly
visible in Fig.~\ref{fig:scaled}.

\begin{figure}[h!]
  \includegraphics[height=.25\textheight]{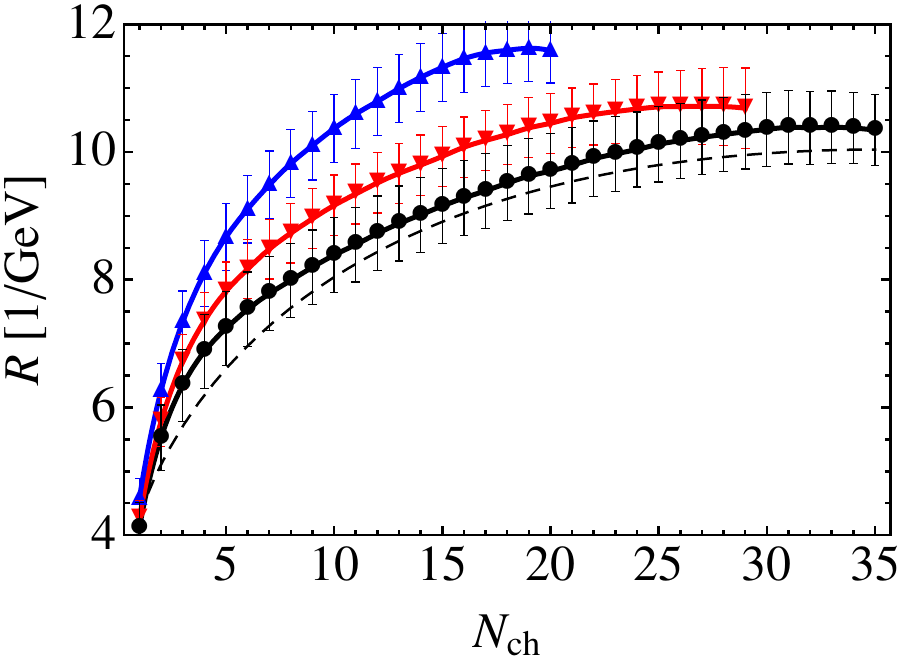}
  \caption{Interaction  radii for pp collisions at 0.9 TeV (blue) up-triangles, 2.76 TeV
  (red) down-triangles and 7 TeV (black) full circles extracted from the 
  mean $p_{\rm T}$ data by means
  on the first equation of formula  (\ref{pTWdep}) with $\beta=1$ and $\alpha=0.285$
  together with experimental errors. Dashed line corresponds to the CGC parameterization
  of Ref.~\cite{McLerran:2014apa}.}
  \label{extractedR}
\end{figure}

To obtain plots in Fig.~\ref{fig:scaled} we have used parametrization of $R(\sqrt[3]{N_g})$ 
at $W_0=7$~TeV from Ref.~\cite{McLerran:2013oju}. It is however clear from Eq.~(\ref{RWdep})
that for energies smaller than $W_0$, but still large, the interaction radius can become
exceedingly large, especially for large multiplicities. We expect therefore that for large 
multiplicities interaction radii at different energies should tend to a common limiting value \cite{McLerran:2014apa}.
The details of this saturation can be extracted from the data. To this end we have used
first equation in formula (\ref{pTWdep}) to extract shape of $R(N_{\rm ch})$ (assuming $\beta=1$).
In order to fix the only remaining parameter $\alpha$ we have required scaling law (\ref{RWdep})
to be valid over the largest possible range of $N_{\rm ch}$ with the result $\alpha=0.285$.
The interaction radii for pp collisions at different energies  obtained in this way
are  plotted in Fig.~\ref{extractedR}. 
We see that indeed they
show a kind of turnover for large multiplicities -- a sign of possible convergence to
the scaling value at $N_{\rm ch} \rightarrow \infty$. This may be verified by future measurements
of mean $p_{\rm T}$. Let us stress, however, that from this point of view, measurements of
$\left\langle p_{\rm {T}}\right\rangle $ for very large $N_{\rm ch}$ at smaller energies would
be even more interesting. On the same plot in Fig.~\ref{extractedR} we also show the CGC
calculation of $R(\sqrt[3]{\gamma N_{\rm ch})}/\beta$ with $\gamma=1.138$ as extracted from
the data in Ref.~\cite{McLerran:2014apa} (dashed line). We see quite good agreement with the model independent
result at 7 TeV represented by (black) full circles.

The  analysis presented here shows that an intriguing insight into the
interaction  geometry in multiple particle production can be gained
by studying mean $p_{\rm T}$ correlation
with $N_{\rm ch}$. A good description of recent ALICE data is obtained when one assumes that
number of particles produced at given impact parameter is proportional to the overlap
{\sl volume} of the interacting hadrons rather than to the {\sl transverse area} alone. This means that  
the evolution time in the longitudinal direction is larger for larger multiplicities. Similar mechanism
has been proposed within the framework of the multipomeron exchange model \cite{Armesto:2008zz}
where one postulates that the string tension of the color system stretched after the collision
 gets larger for larger multiplicities. In our analysis based on saturation and geometrical scaling fixed
 number of particles is produced from smaller volume when the collision energy increases. This means
 that when multiplicity is fixed collisions are more peripheral at high energy and more central at
 lower energies. However, since there is a natural cutoff on the size of the interaction volume corresponding
 to the size of interacting hadrons, one expects that for very large multiplicities interaction radii tend to
 the fixed limit independently of energy. This trend seems to be seen already in the existing LHC
 data. Future data will shed more light on this interesting issue.

%%%%%%%%%%%%%%%%%%%%%%%%%%%%%%%%%%%%%%%%%%%%%%%%
%% BACKMATTER
%%%%%%%%%%%%%%%%%%%%%%%%%%%%%%%%%%%%%%%%%%%%%%%%

\begin{theacknowledgments}
Many thanks to the organizers of the 
International Workshop on Diffraction in High-Energy Physics 2014, Primo{\u{s}}ten, Croatia,
for the stimulating and interesting meeting.
The research of M.P. has been supported by the Polish NCN grant 2011/01/B/ST2/00492.  
%The research of L.M. is supported under DOE Contract No. DE-AC02-98CH10886. 
\end{theacknowledgments}



\bibliographystyle{aipproc}   % if natbib is available

%\end{document}

%%%%%%%%%%%%%%%%%%%%%%%%%%%%%%%%%%%%%%%%%%%
%% The following lines show an example how to produce a bibliography
%% without the help of the BibTeX program. This could be used instead
%% of the above.
%%%%%%%%%%%%%%%%%%%%%%%%%%%%%%%%%%%%%%%%%%%

%\endinput
\end{document}